\documentclass[conference]{IEEEtran}
\ifCLASSINFOpdf
  \usepackage[pdftex]{graphicx}
  \graphicspath{{../pdf/}{../jpeg/}}
  \DeclareGraphicsExtensions{.pdf,.jpeg,.png}
\else
   \usepackage[dvips]{graphicx}
   \DeclareGraphicsExtensions{.eps}
\fi
\begin{document}
\title{A Bayesian approach to the inference of parametric configuration of the signal-to-noise ratio in an adaptive refinement of the measurements}

\author{\IEEEauthorblockN{Author: Maria Jose Marquez}
\IEEEauthorblockA{UNED University \\Artificial Intelligence\\
Madrid\\
Email:mmarquez92@alumno.uned.es\\
}
}


%Telephone: (800) 555--1212, Fax: (888) 555--1212}


\maketitle

\begin{abstract}
%\boldmath
Calibration is nowadays one of the most important processes involved in the extraction of valuable data from measurements. The current availability of an optimum data cube measured from a heterogeneous set of instruments and surveys relies on a systematic and robust approach in the corresponding measurement analysis. In that sense, the inference of configurable instrument parameters can considerably increase the quality of the data obtained.\\
Any measurement devoted to scientific purposes contains an element of uncertainty. The level of noise, for example, determines the limit of usability of an image. Therefore, a mathematical model representing the reality of the measured data should also include at least the sources of noise which are the most relevant ones for the context of that measurement.\\
This paper proposes a solution based on Bayesian inference for the estimation of the configurable parameters relevant to the signal to noise ratio. The information obtained by the resolution of this problem can be handled in a very useful way if it is considered as part of an adaptive loop for the overall measurement strategy, in such a way that the outcome of this parametric inference leads to an increase in the knowledge of a model comparison problem in the context of the measurement interpretation.\\
The context of this problem is the multi-wavelength measurements coming from diverse cosmological surveys and obtained with various telescope instruments. As a first step,, a thorough analysis of the typical noise contributions will be performed based on the current state-of-the-art of modern telescope instruments,  a second step will then consist of identifying configurable parameters relevant to the noise model under consideration, for a generic context of measurement chosen. Then as a third step a Bayesian inference for these parameters estimation will be applied, taking into account a proper identification of the nuisance parameters and the adequate selection of a prior probability. Finally, a corresponding set of conclusions from the results of the implementation of the method proposed here will be derived 
\end{abstract}


% \fi
\IEEEpeerreviewmaketitle

\section{Introduction}
\label{sect:1}
As indicated in \cite{AstronomicalPhotometry:Budding}, astronomical photometry is about the measurement of the brightness of radiating objects in the sky. Many factors, such as those coming from the instrument limitation, from a fixed measurement strategy, or the limitation from the mean through which the measurement is taking place, make this area of the science relatively imprecise. The improvement in the dectectors technology plays a key role in the area of optimizing the resulting astronomical photometric measurements. In this sense, a signal-to-noise ratio capable of being configured as part of an optimization framework of the measurement system seems to be a useful input.\\
Charge-coupled devices (CCDs) constitutes the state-of-the-art of detectors in many observational fields. \cite{CCDHandbook:Howell} enumerates the areas involved in the recent advances of the CCDs systems, which are:
\begin{itemize}
 \item 
Manufacturing standards that provide higher tolerances in the CCD process leading directly to a reduction in their noise output.
\item
Increased quantum efficiency, especially in the far red spectral regions.
\item
New generation controll electronics with the ability for faster readout, low noise performance, and more complex control functions.
\item
New types of scientific grade CCDs with some special properties.
\end{itemize}

Any data, in general, is always limited in accuracy and incomplete, therefore, deductive reasoning does not seem to be the proper way to prove a theory. However, and  as said in \cite{BayesianPhysicalSciences:Gregory}, statistical inference provides a mean for estimating the model parameters and their uncertainties, which is known as data analysis. It also allows assessing the plausability of one or more competing models.

The use of a Bayesian approach here is also justified in \cite{BayesianPhysicalSciences:Gregory} where it is stated that for data with a high signal-to-noise ratio for example, a Bayesian analysis can frequently yield many orders of magnitude improvement in model parameter estimation, through the incorporation of relevant prior. This is exactly what we intend through the implementation of what will be described in this paper, and detailed in the following section.

\section {Description of the problem to be resolved} 
\label{sect:2}
The problem to be resolved here consists in the implementation of bayesian inference for a set of configurable parameters which affect the signal-to-noise ratio of a measurement. In comparisson with other methodologies, such as ANOVA, which shows serious weakness when outliers are present in the measured data, bayesian parameter inference offers a robust method against outliers. It also allow to improve the results of inference by using the posterior probability density distribution (pdf) of one execution as the prior pdf for another execution in a recursive framework. This will lead to an adaptive measurement strategy which can be addressed as a calibration refinement.

Professional  surveys plan the measurement strategy well in advance, taking into account all the relevant factors impacting on the measurement; this involves the set of specified fix parameters from the detector and also a set of parameters which configure the measurement, such as integration time, diameter of the aperture, etc.

Once a measurement has finished, the data are archived and their analysis and processing begin. The problem proposed here is to establish a link between the results of a measurement under a specific detector configuration and the refinement, by application of parameter bayesian inference,  of the configuration parameters to be applied in a further measurement. The result of this bayesian inference at parameter level is proposed to be injected as additional knowledge for a model selection problem in the context of measurement data analysi (i.e, photometric cross-matching of multiwavelength astronomical sources).

For example, let us imagine that we have performed colour measurement in a multi-wavelength survey  with ten different instruments, each one under a specific configuration. Let us imagine that in the process of model comparison for the cross-matching scenarios, the existence of a source inferred in a bandwith which is not detecting it, is plausible. Then, based on this result, a new configuration for that instrument can be inferred in such a way that allows us to explore the refined plausability indicated by the model comparison from the data obtained in the first measurement loop.
Figure \ref{fig1} shows a  block diagram reflect in general lines the idea of the problem proposed here\\

\begin{figure}[Graficobis]
\label{fig1}
\centering
\includegraphics[width=2.5in]{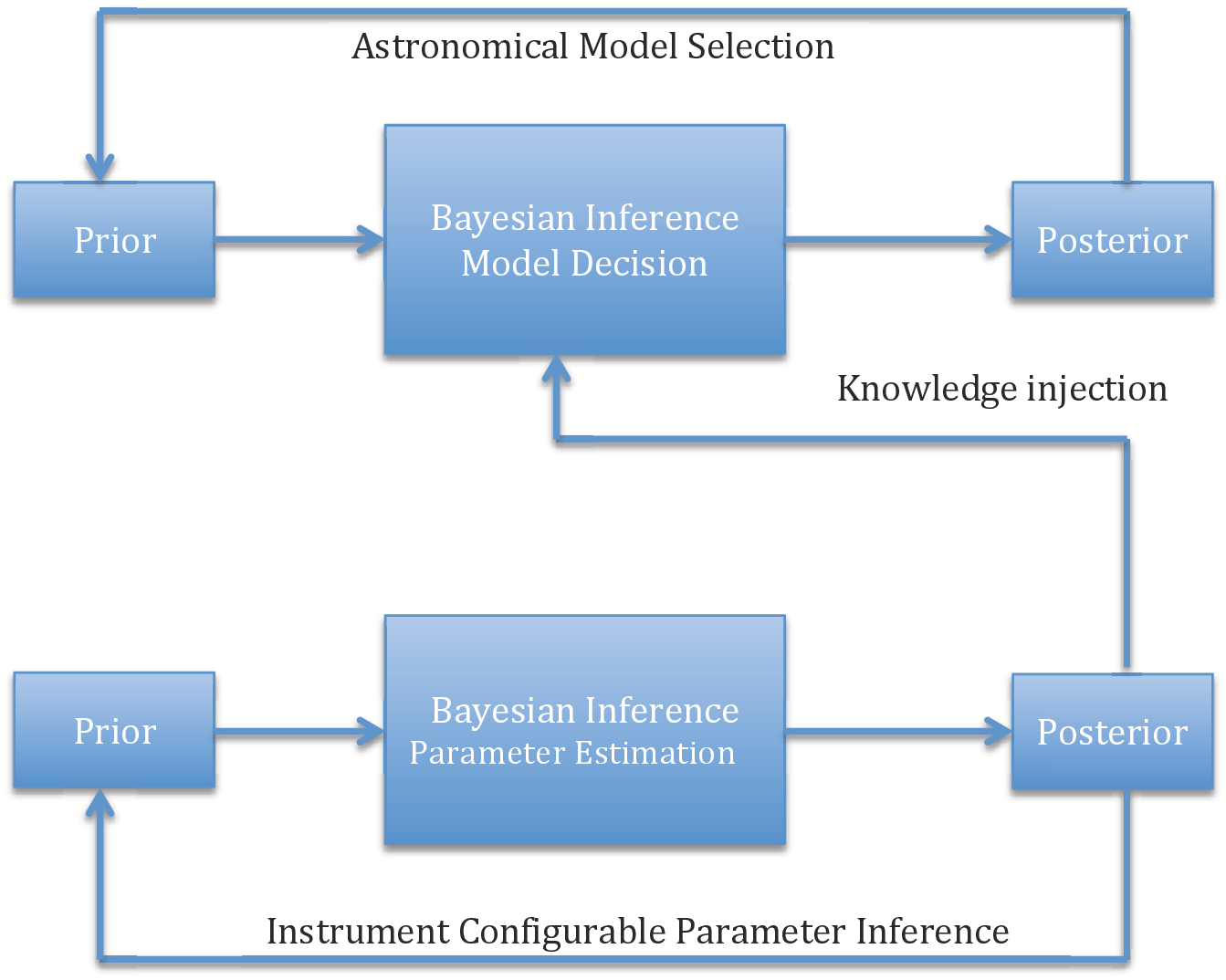}
% where an .eps filename suffix will be assumed under latex, 
% and a .pdf suffix will be assumed for pdflatex; or what has been declared
% via \DeclareGraphicsExtensions.
\caption{Two levels of Bayesian Inference}
\label{fig1}
\end{figure}

\section{State of the art of telescope detectors}
\label{sect:3}
The great majority of detectors used in the astronomical field are Silicon-based ones; this means that the electronics involved in the specification, manufacturing and operational life of these detectors are relvant to the outcome obtained. As detailed in \cite{AstronomicalPhotometry:Budding}, the excitations of electrons responding to incident photons constitutes the fundamentals for practical flux measurement in almost  all nowadays photometric systems.\\
In nowadays, CCDs are used in many instruments involved in the main current astronomical surveys; an extensive and increasing bibliography is currently being publishing. Therefore our paper will focussed on this type of instruments, however we have tried to retain the generic aspect of the characterization of any other type of detector and obviously the methology presented here is fully valid with any other set of specific parameters.
\subsection{Characterization of telescope detectors}
A summary from \cite{AstronomicalPhotometry:Budding}on the relevant information related to detector parameters has been included in this subsection for the sake of clarity.
A more detailed information can be found in \cite{AstronomicalPhotometry:Budding}\\
In general terms and for the context of our problem, a detector can be characterized by the following parameters, as indicated in \cite{AstronomicalPhotometry:Budding}:\\
\begin{itemize}
 \item 
\textbf{Quantum Efficiency (Q)}: it is the ratio between registered events and the incident photon. 
\item
\textbf{Information Transfer Efficiency (E)}: it is the ration between the square of the signal-to-noise ratio of the output and the square of the signal-to-noise ratio of the input.\\
\begin{equation}
\label{eq1}
 E = \frac{(S/N)_{out}^2}{(S/N)_{in}^2}
\end{equation}
\item
\textbf{Noise equivalent power (NEP)}:it is defined as the optical power producing an output equal to the noise level of the detector.
\begin{equation}
\label{eq2}
NEP=\frac{fA}{(S/N)(\Delta\nu)^{(1/2)}}
\end{equation}
\item
\textbf{Linearity and Saturation}: one of the most relevant characteristics of a CCD detector is its linear property. This means that, under ideal scenario (ignoring effects such as, noise, dark current, polarization current, sky background contribution etc), the intentsity registered in each pixel (as elecrons) is proportional to the incident light. However this linear behaviour has its limits. The most obvious one is the \textbf{Saturation threshold}, which is measured by the \textbf{Full-well capacity}
\item
\textbf{Full-well capacity}:this parameter measures the limit of the accumulated charge before saturation begins. This value is normally included in the technical specification of the detector's supplier.
\item
\textbf{Event or pixel capacity}: it is defined as how many events can be usually accumulated before some saturation effect takes place.
\item
\textbf{Working range}: also named \textbf{Dynamic range}, is essentially the same as event capacity for a noiseless accumulative single detector, but more generally interpreted as the difference between useful maximum and minimum event counts.
\item
\textbf{Gain(G)}: the digital image consists of a table of numbers which indicate the intensity registered in each pixel. However, the numbers stored do not mean the quantity of electrons found in each electrode, as this quantity can be huge and this would make the resultant storage files too big. Therefore what is normally done is to divide the quantity of electrons by a certain number, named \textbf{Gain}; thus, what we register in the file is the number of \textbf{counts} obtained when performed the above division. Sometimes counts are called \textbf{ADU} (Analog-to-Digital Units) or \textbf{DN} (Data Number). The Gain is therefore measured in electrons per count.\\
Some cameras allow the user to choose the Gain. Then we could choose a small value for faint detections or bit to measure correctly sources of various brightness, but all this without overpassing the limit done by:\\
\begin{equation}
\label{eq3}
Maximum Gain = \frac{pixel capacity}{dynamic range}
\end{equation}
\item
\textbf{Counts(c)}: The number of photons that fall on a pixel is related to the counts by:\\
\begin{equation}
\label{eq4}
Photons =  \frac{G . c}{Q} 
\end{equation}
\end{itemize}

Detector parameters such as quantum efficiency, linearity event capacity and so on, are often characterized by \textbf{figures of merit}, which manufacturers quote about their products.

\subsection{Review of the main sources of noise in the photometry measurement}
The error which all scientific measurements should carry means really uncertainty and it is due to noise. Following the line of discussion presented at the begining of this section, once an electron has been excited by a photon, the next step consists in registering this event by the electronics of the detector. In this process, handling and reducing the number of extraneous electron activity which does not come from the source subject to the pure measurement (noise) consitutes a delicate and complex step of this process. A summary from \cite{AstronomicalPhotometry:Budding}on the relevant information related to sources of noise has been included in this section for the sake of clarity.
For more detailed information about the functional block description of any detector can be found in \cite{AstronomicalPhotometry:Budding}\\
There are numerous sources of noises in the CCD images. The following list identify those sources of noise which are more relevant to the problem described  in section \ref{sect:2}:
\begin{itemize}
 \item 
\textbf{Dark current}; also named \textbf{Thermal noise}, it is produced by the spontaneous generation of electrons in the Silicium (Si) due to the themal excitation of the material.\\
The noise associated with dark current in a CCD is regarded as having a spatial dependence, in that it relates to minor irregularities in the solid state molecular bonding lattice, associated with material nterfaces and impurities. Each pixel generates a slightly different level of dark current, so the noise depends on non-uniformity of the response over the surface as well as the dark current's inherent quasi-Poissonian contribution to the electron counts. Dark current, following a generally Richardson law type dependence on temperature, could fill the potential wells of an uncooled CCD at $20\,^{\circ}{\rm C}$ in, typically, a few seconds, but cooling to $-50\,^{\circ}{\rm C}$ say, reduces this to a tolerable few electrons per pixel per second. The standards approach to spatial non-uniformity of response in CCD is through \textbf{flat fielding}, although it should be rememberd that there is an additional shot noise contribution to the adopted flat field contribution.\\
Within the CCD, each pixel has a slightly different gain or quantum efficiency when compared with its neighbors. In order to flatten the relative response for each pixel to the incoming radiation, a flat field image is obtained and used to perform this calibration. Ideally, a flat field image would consist of uniform illumination of every pixel by a light of source of identical spectral response to that of the object frames. Once a flat field image is obtained, one them simply divides each object frame by it implementing then instant removal of pixel-to-pixel variations.
\item
\textbf{Cosmic rays}: they are part of the inhabitants of the interstellar space; cosmic rays are sub-atomic particles which go at very high speed, near to the speed of light. Cosmic rays are very annoying for the astronomers because they interact with the Silicium of the coupled charged devices. These electronic micro-aluds, concentrated in a few pixels appear in the images as points, or sometimes, lines very bright. As much the exposition time is, more quantity of cosmic rays will deteriorate it.
\item
\textbf{Read noise}; this is a very important contribution to the toal noise of the images. This noise is due to the random and unavoidable errors that are produced during the reading of the image, in the process of amplification and counting of the electrons captured in each pixel. These errors are intrinsic to the nature of the detector device.\\
The existance of read noise has always to be taken into account because it affects to all steps towards the obtainment and traitment of the digital images. Each camera must have its own level of read noise, documented in its technical specifications.
\item
\textbf{Shot noise}: this is a statistical noise due to the inherently non-steady photon influx. A Poissonian distribution is normal for the arrival of the primary photon stream from a source of constant emissive power.
\item
\textbf{Clocking noise}; this noise comes from the various high frequency oscillators involvd in the gating circuitry. This noise rises with load and clocking frequency, but it can normally be controlled by manufacturers to a negligible level for astronomical applications.
\item
\textbf{Atmosphere}: The total noise of detection is not just that of the photoelectric effect on the detector, since the signal has already been deteriorated by the atmosphere. Further reading on the problems with atmosphere in the astronomical measurements can be found in \cite{IntroductiontoCCD:Romanishin}
\end{itemize}

For simplification purposes in the resolution of the problem describe in \ref{sect:2}, only the following list of noise sources will  be considered representative for the context of our problem. However the methodology can be extended to a more exhaustive list of noises: dark current, read noise, background noise.

\section{Signal-to-Noise ratio and configurable parameters}
As detailed in \cite{CCDHandbook:Howell}, a careful understanding of the main sources of uncertainties can suggest ways to improve our measurement strategy, this means, the observation, reduction and analysis processes.\\
A crucial concept in photometry is the \textbf{signal-to-noise ratio (S/N)}, which is equivalent to the concept of percentage error $(N/S)\times(100\%)$\\
It is very important to assess and, if possible, to reduce the noise of the images.In this direction, the parameter $S/N$ is very useful in the assessment of the feasibility, reliability and quality of the detection.
As a general rule, and based on the considerations expressed in \cite{ManualPracticoCCD:Galadi}, to get reliable photometry and/or astrometry measurements, the minimum threshold must be: $S/N = 4$\\
As a first preliminary simplification, under the asumption of photon noise dominating the noise, the counting statistics of the number of photons impacting on a given area per second can be modelled by a Gaussian distribution, where the scatter is the square root of the number of photons, therefore:
\begin{equation}
\label{eq5}
S/N = \frac{n}{\sqrt{n}}
\end{equation}
Where $n$ includes the photon counts for the sky foreground and the sky background, both of them carrying noise components. To obtain the count from the source alone, the sky background contribution has to be substracted. Then, considering this two contributions, we can write the following:
\begin{equation}
S/N = \frac{C_{source}}{\sqrt{C_{source} + 2\times C_{background}}}
\end{equation}
The S/N ratio changes with the integration time, $t$ and with the telescope aperture,$D_{tel}$, therefore these two parameters encompass the configuration domain by which the S/N ratio can be optimized in the process explained in section \ref{sect:2}. \\
For the telescope aperture, we know that increasing the diameter of the telescope primary,$D_{tel}$, by a factor of 2 increases the collecting area by  factor of 4, thus for a given integration time, $t$ we get: $ (S/N)  \alpha   D_{tel}$\\
Regarding the integration time dependency with the signal-to-noise ratio, we can write:
\begin{equation}
\label{eq6}
S/N = \frac{tR_{star}}{\sqrt{tR_{star}+2tR_{sky}}}
\end{equation}
Where $R$ is th count rate expressed in $countss^{-1}$. Therefore, $S/N \alpha \sqrt{t}$

Similarly the number of exposures impact on the noise of the resulting image. According to \cite{ManualPracticoCCD:Galadi} if the exposure is broken into $n$ equal short exposures, the error in the mean of the measurements would be:
\begin{equation}
\label{eq7}
\sigma_{mean} = \frac{\sigma_{individ}}{\sqrt{n_{meas}}}
\end{equation}
Where $\sigma_{individ}$ is the scatter in the individual short exposures and $n_{meas}$ is the number of such short exposures.It is important to keep in mind that if we add $n$ images resulting from $n$ short exposures, the resulting signal is $S=s_1+s_2+\dots+s_n$, and total noise of the resulting image will be $R = \sqrt{r_1^2+r_2^2+\dots+r_n^2}$
As a conclusion, the signal-to-noise ratio of an addition of images from the same object is bigger and therefore better than the signal-to-noise ration of each individual image.

So far, photon noise has been assumed to be the dominant contribution of the noise, and therefore the other noise sources contributions have been reduced to zero for the $S/N$ equations above. However, in the case of faint sources detection  the dark current and the read noise can play a key role in the $S/N$, therefore, we will develop the $S/N$ equation, with the integration time dependency and including at least the following sources of noise: photon noise, dark current and read noise. For each of the noise sources, valid approximations will be considered in order to obtain a final $S/N$ equation which is computational cost affordable for a intel mac Core 2 Duo computer.

Therefore, the equation for the $S/N$ of a measurement made with a CCD can be given by:
\begin{equation}
\label{eq8}
S/N = \frac{N_*}{\sqrt{N_* + n_{pix}(N_S + N_D + N_R^2)}}
\end{equation}
Where:
\begin{itemize}
 \item 
$N_*$ is the total number of photons which compound the signal detected from the object
\item
$N_S$ is the total number of photons coming from the background also called sky
\item
$N_D$ is the total number of dark current electrons per pixel
\item
$N_R^2$ is the total number of electrons per pixel resulting from the read noise.
\end{itemize}

The noise terms in equation \ref{eq8} can be modelled by Poisson distributions. The term $n_{pix}$ is used to apply each noise term on a per pixel basis to all the pixels involved in the $S/N$ measurement. A more complete equation taking into account digitization noise within the A/D  converter can be found in \cite{CCDHandbook:Howell} \\

As explained in \cite{CCDHandbook:Howell}, and using the fact that $S/N = 1/\sigma$, a standard error for the measurement can be obtained as:
\begin{equation}
\sigma_{magnitude} = \frac{1.0857\sqrt{N_*+p}}{N_*}
\end{equation}

where $p$ is equal to the noise terms indicated in \ref{eq8}, and $1.0857$ is the correction term between an error in flux (electrons) and that same error in magnitudes (Howell, 1993).

The equation \ref{eq8} can also be expressed in terms of count rate and integration time, as follows:
\begin{equation}
\label{eq9}
S/N = \frac{Nt}{\sqrt{Nt + n_{pix}(N_{S}t+ N_{D}t + N_{R}^2)}}
\end{equation}

And in line with the text above, each terms of equation \ref{eq8} can be expressed as follows:
\begin{equation}
\label{eq10}
N ={C_1}\cdot  \frac{G\cdot c}{Q}
\end{equation}

\begin{equation}
\label{eq11}
N_{S}=C2\cdot \sqrt{2C_{sky}}
\end{equation}

\begin{eqnarray}
\label{eq12}
N_{D}=C3\cdot 5.86\cdot10^{9}\cdot T^{\frac{2}{8}}\cdot exp{\frac{-E_g}{2KT}} \\
E_g = 1.1557- \frac{7.021\cdot 10^{-4}\cdot T^2}{1108+T}
\end{eqnarray}

Where $C_1$,$C_2$ and $C_3$ are proporcionality constants.

\section {Bayesian inference: estimation of the integration time}
In general, as described in \cite{BayesianPhysicalSciences:Gregory}, a Bayesian Probability Density Function is a measure of our state of knowlege of the value of the parameter. When we acquire some new data, Bayes' theorem provides a means for combining the information about the parameter coming from the data, through the likelihood function, with the prior probability, to arrive at a posterior probability density, $p(H|D,I)$, for the parameter.\\

Let us be $M$ the known model for the signal-to-noise ratio of a multi-wavelength measurement system which identifies cross-matching sources, ${\Phi}$ the set of configurable parameters,$\{t,D_{tel}\}$, being $t$ the integration time and $D_{tel}$ the configurable aperture diameter of the instrument, and ${D}$ the data of the measurement. $I$ is the information associated to the model. By application of Bayesian inference, the configurable parameters for a established model $M$ can be estimated as follows:

\begin{equation}
\label{eq13}
p(\Phi|D,M,I) = \frac{p(\Phi|M,I).p(D|M,\Phi,I)}{p(D|M,I)} 
\end{equation}

This equation can be written in the following way:
\begin{equation}
\label{eq14}
 p(\Phi|D,M,I)=\frac{p(\Phi|M,I).p(D|M,\Phi,I)}{\int_{-\infty}^{+\infty}p(\Phi|M,I).p(D|M,\Phi,I)d\Phi}
\end{equation}

Therefore it becomes evident that the denominator of \ref{eq2} is just a normalization factor and we can focus on just the numerator, where $p(\Phi|M,I)$ is named the probability a priori and $p(D|M,\Phi,I)$ is the likelihood.

Strictly speaking, as it is explained in \cite{BayesianPhysicalSciences:Gregory}, Bayesian inference does not provide estimates for parameters; rather, the Bayesian solution to the parameter estimation problem is the full posterior PDF, $p(\Phi|D,M,I)$ and not just a single point in the parameter space. It is useful to summarize this distribution and one possible candidate of the best-fit value is the posterior mean, 

\begin{equation}
 \label{eq15}
<\Phi> = \int \Phi p(\Phi|D,M) d\Phi
\end{equation}

We will develop here the bayesian estimation parameter for the integration time $t$, however the consideration of more than one parameter is immediate and in that case multiple integrals will be considered for the marginalization of the corresponding nuisance parameters.

A preliminary choice for the prior probability will be a Uniform distribution, therefore:
\begin{equation}
\label{eq16}
p(t|M,I) = \frac{1}{\Delta t}= \frac{1}{t_{max}-t_{min}}
\end{equation}
where $t_{max}$  and $t_{min}$ are the maximum and minimum integration time to be defined for the observation under consideration.

In general, the difference between the data and the model is called the error, therefore:
\begin{eqnarray}
\label{eq17}
p(D|M,t,I) = p(D_1,...,D_N|M,t,I) \\ \nonumber
= p(E_1,...,E_N|M,t,I) \\ \nonumber
= \prod_{i=1}^N p(E_i|M,t,I)
\end{eqnarray}

The model here $M$ is the one proposed in \ref{eq9},however for the sake of clarity in the following expressions, we will consider that equation \ref{eq9} can be simplified as follows:

\begin{equation}
\label{eq18}
S/N = C \cdot \sqrt{t}
\end{equation}
Being $C$ a constant

The likelihood for all the channels $N$ under consideration in a multiwavelength observation, can be expressed as follows:
\begin{equation}
\label{eq19}
p(D|t,M,I) = \prod_{i=1}^{N} \frac{1}{\sigma\sqrt{2\pi}}exp{\frac{-(d_i - C\sqrt{t})^2}{2\sigma^2}}
\end{equation}

Now we can compute $p(t|D,M,I)$ as indicated in \ref{eq14}, and considering a normalizing unity constant for the denominator we obtain the following result:.
\begin{eqnarray}
 \label{eq20}
p(t|D,M,I) = p(t|M,I)p(D|M,t,I) \\ \nonumber
= \frac{1}{\Delta{t}}(2\pi)^{-N/2}\sigma^{-N}exp\{\frac{-\sum_i (d_i-C\sqrt{t})^2}{2\sigma^2}\} \\ \nonumber
= \frac{1}{\Delta{t}}(2\pi)^{-N/2}\sigma^{-N}exp\{\frac{-\sum_i d_i^2}{2\sigma^2}\}exp\{\frac{-Nt}{2\sigma^2}\}exp\{\frac{\sum_i(2d_i\sqrt{t})}{2\sigma^2}\}
\end{eqnarray}

\section {Conclusion}
The main purpose of the methodology presented here is to enable the capability of an observational system to adapt its configurable parameters depending on the results from previous observations. In this way, It is clear that the integration time has a considerably strong impact on the faint sources detections, therefore the approach covered in the previous chapter of this paper intends to infer the optimum integration time in order to increase the probability of detection for the next observation to be planned and defined. The inclusion in the model of the main contributions of noise leads to a more refined parametric inference.

A toy example has been built in order to show some preliminary results derived from the approach presented in this paper. Figure \ref{fig2} represents the data in terms of integration times from several observations and Figure \ref{fig3} shows the probability density corresponding to the posterior probability solved as detailed in the previous section. 

A real example is planned to be implemented using a model of signal-to-noise ratio in line with the expression presented in equation \ref{eq20}

\begin{figure}[Ejemplo]
\label{fig1}
\centering
\includegraphics[width=2.5in]{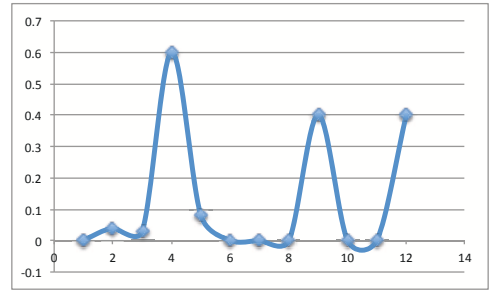}
% where an .eps filename suffix will be assumed under latex, 
% and a .pdf suffix will be assumed for pdflatex; or what has been declared
% via \DeclareGraphicsExtensions.
\caption{Toy example: integration time from different observations}
\label{fig1}
\end{figure}

\begin{figure}[Resultado]
\label{fig1}
\centering
\includegraphics[width=2.5in]{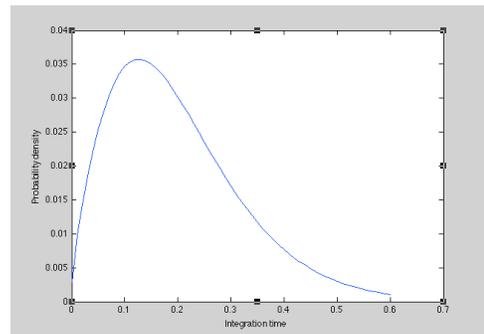}
% where an .eps filename suffix will be assumed under latex, 
% and a .pdf suffix will be assumed for pdflatex; or what has been declared
% via \DeclareGraphicsExtensions.
\caption{Probability density corresponding to the integration time inference}
\label{fig1}
\end{figure}

\section*{Acknowledgment}

The authors would like to thank Dr. Luis Manuel Sarro Baro and Dr. Tamas Budavari for his support and valuable insights in the area of bayesian inference, and Mr. Georges Bernede for his continuous support in the effective conciliation of  my professional and academic activities.



%\bibliographystyle{IEEEtran}

\begin{thebibliography}{1}

\bibitem{IEEEhowto:kopka}
H.~Kopka and P.~W. Daly, \emph{A Guide to \LaTeX}, 3rd~ed.\hskip 1em plus
  0.5em minus 0.4em\relax Harlow, England: Addison-Wesley, 1999.
\bibitem{CCDHandbook:Howell}
Steve B. Howell, \emph{Handbook of CCD Astronomy}, 2nd~ed.\hskip 1em plus
  0.5em minus 0.4em\relax National Optical Astronomy Observatory and WIYN Observatory: Cambridge University Press.
\bibitem{AstronomicalPhotometry:Budding}
Edwin Budding and osman Demircan, \emph{Introduction to Astronomical Photometry}, 2nd~ed.\hskip 1em plus
  0.5em minus 0.4em\relax Canakkale University, Turkey: Cambridge University Press.
\bibitem{IntroductiontoCCD:Romanishin}
W. Romanishin, \emph{An Introduction to Astronomical Photometry Using CCDs}, September 16, 2000.\hskip 1em plus
  0.5em minus 0.4em\relax University of Oklahoma.
\bibitem{ManualPracticoCCD:Galadi}
David Galadi Enriquez and Ignasi Ribas Canudas, \emph{Manual Practico de Astronomia con CCD}, 1998.\hskip 1em plus
  0.5em minus 0.4em\relax Department of Astronomy and Metrology of University of Barcelona: Omega.
\bibitem{StatisticsAstronomers:Wall}
J. V. Wall and C. R. Jenkins, \emph{Practical Statistics for Astronomers}, 2003.\hskip 1em plus
  0.5em minus 0.4em\relax University of Oxford and Schlumberger Cambridge Research: Cambridge University Press.
\bibitem{BayesianMethods:Trotta}
Michael P .Hobson,  Andrew H. Jaffe, Andrew R. Liddle, Pia Mukherjee and David Parkinson, \emph{Bayesian Methods in Cosmology}, 2010.\hskip 1em plus
  0.5em minus 0.4em\relax Cavendish Laboratory, University of Cambridge, Imperial College, London, University of Sussex: Cambridge University Press.
\bibitem{BayesianPhysicalSciences:Gregory}
Phil Gregory, \emph{Bayesian Logical Data Analysis for the Physical Sciences}, 2005.\hskip 1em plus
  0.5em minus 0.4em\relax Department of Physics and Astronomy, University of British Columbia: Cambridge University Press
\end{thebibliography}

\end{document}